\documentclass[preprint,aps,draft]{revtex4}

\usepackage{bm}

\begin{document}


\title{Possibility of automatic Pauli-Villars-like regularization through 
extra dimensions}

\author{Recai Erdem}
\email{recaierdem@iyte.edu.tr}
\affiliation{Department of Physics,
{\.{I}}zmir Institute of Technology \\ 
G{\"{u}}lbah{\c{c}}e K{\"{o}}y{\"{u}}, Urla, {\.{I}}zmir 35430, 
Turkey} 

\date{\today}

\begin{abstract}
Fermions in a space with some extra dimensional reflection
symmetries are considered. In this space only one Kaluza-Klein mode is 
observed at the scales larger than the sizes of the extra dimensions 
while at the smaller scales all modes become observable. The resulting 
picture suggests that this model has a built-in Pauli-Villars-like 
regularization as its inherent characteristic. 
\end{abstract}

\pacs{11.10.Kk,12.60.-i,11.10.Gh}
\maketitle

\section{Introduction}
Extra dimensions are attractive frameworks to address many problems
in high energy physics. They have a good 
prospect to account for many theoretical and phenomenological problems 
that the standard model can not answer, such as the hierarchy between the 
strengths of gravitational and electroweak interactions, fermion masses, 
families, chirality, cosmological constant problem, Higgs-Gauge 
unification etc. in addition to being the standard setting for 
string theory. The standard practice for extra dimensions is to take 
them be compact (at least at the energies much smaller than the Planck 
scale). This, in turn, results in an infinite number of Fourier modes 
for an extra dimensional field that are called Kaluza-Klein (KK) modes 
of that field \cite{Kaluza-Klein,Duff}. Although extra dimensional models 
are promising candidates for the physics beyond the standard model these 
models have a technically unpleasing aspect; the infinite number of KK 
modes are additional sources for infinities in the corresponding quantum 
field theories. Kaluza-Klein modes make the issue of the regularization 
more intricate even when one lets extra dimensional models be effective 
field theories \cite{Alvarez, Baum-Dienes}. In this paper I introduce a 
model where the Kaluza-Klein modes serve for regularization on contrary to 
the generic case where Kaluza-Klein modes make the regularization more 
difficult.

In some of my recent studies I had considered metric reversal symmetry as 
a possible cure to cosmological constant and zero point energy problems of 
quantum fields \cite{Erdem-cc1,Erdem-cc2}, and in \cite{Erdem-kk} I had 
considered fermions and a variant of the metric reversal symmetry to get 
a finite number of Kaluza-Klein (KK) modes 
and 
their ghosts at the scales larger than the size of extra dimensions while 
all KK modes are observed at smaller scales. In this study I consider 
fermions and use the same symmetry used in \cite{Erdem-kk} to construct a 
space where only one KK mode is observed at the scales larger than the 
size of the extra dimensions while a usual KK mode and its ghost are 
observed at smaller length scales. The resulting picture amounts to an 
automatic built-in Pauli-Villars \cite{Pauli-Villars} like 
regularization. 
The main elements of this scheme are two extra 
dimensional discrete symmetries in conjunction with non-trivial boundary 
conditions imposing specific forms for the Lagrangians at different energy 
scales. The details of the scheme are given in the following sections. In 
the next section the space, the symmetries, and the boundary conditions 
are specified. In the remaining sections the scheme is introduced and 
studied.

\section{The setting, the symmetries, and the boundary conditions of the 
model} 
Consider the following 7-dimensional space
\begin{equation}
ds^2\;=\;g_{\mu\nu}(x)\,dx^\mu 
dx^\nu\,-\,
\cos^2{k_2y_2}\,[\,
\,dy_1^2\,+\,\cos^2{k_3y_3}
dy_2^2\,+\,dy_3^2\,]
~~~~~~~~\mu,\nu\,=\,0,1,2,3
\label{a1} 
\end{equation}
In fact seven is the minimum number of dimensions that can be taken in 
this scheme. This point will be discussed at the end of the paragraph 
after Eq.(\ref{a9}). I take the extra dimensions be compact and have the 
sizes $L_1$, $L_2$, $L_3$, and $k_1\,=\,\frac{2\pi}{L_1}$, 
$k_2\,=\,\frac{2\pi}{L_2}$, $k_3\,=\,\frac{2\pi}{L_3}$. 
The action for matter fields in this space is 
\begin{equation} 
S_{f} \;=\; \int {\cal L}_{f}\,
\cos^3{k_2y_2}\,
\cos{k_3y_3}\,
d^4x\,dy_1\,dy_2\,dy_3
\label{a2}
\end{equation}
where ${\cal L}_f$ denotes the Lagrangian corresponding to matter fields. 
Note that the extra dimensional contribution to the Einstein-Hilbert 
action for the metric (\ref{a1}) vanishes after integration over $y_2$ 
and $y_3$. In other words (\ref{a1}) is effectively equivalent to its 
4-dimensional part at the scales much larger than $L_2$ and $L_3$. So one 
does not need to bother with energy-momentum tensor content necessary to 
support the extra dimensional piece of (\ref{a1}) at current 
accessible scales. 

I consider the following 
transformations \begin{eqnarray} &&x^a\;\rightarrow\;-\,x^a~~,~~~~
a=0,1,2,3,5 \label{a3a} \\
&&x^b\;\rightarrow\;-\,x^b
~~,~~~~
b=0,1,2,3,6 
\label{a3b}
\end{eqnarray}
where $x^5=y_1$, $x^6=y_2$. 
The general Fourier decomposition 
of a field $\varphi$ in the coordinate $z$ in the presence of the 
symmetry (\ref{a3a}) or (\ref{a3b}) in either of the 5th or 6th directions 
in space may be expressed as
\begin{eqnarray}
\varphi(x,z)&=&\sum_{n=-\infty}^{\infty}\,
[\,\alpha_n(x)\,\,\sin{(\frac{1}{2}n\,kz)}\,
+\,\beta_n(x)\,\,\cos{(\frac{1}{2}n\,kz)}\,] 
\nonumber \\
&=&\sum_{n=0}^{\infty}
\varphi_{|n|}(x)\,\,\sin{(\frac{1}{2}|n|\,kz)}\,
+\,\tilde{\varphi}_{|n|}(x)\,\,\cos{(\frac{1}{2}|n|\,kz)}\
\nonumber \\
&=&\sum_{n=0}^{\infty}
\,[\,a_{|n|}\,\sin{(\frac{1}{2}|n|\,kz)}\,
+\,b_{|n|}\,\cos{(\frac{1}{2}|n|\,kz)}\,]\,\varphi_{|n|}(x)
\nonumber \\
&=&\sum_{n=0}^{\infty}
\{\,f_{|n|}[
\cos{(\frac{|n|kz}{2})}+\sin{(\frac{|n|kz)}{2})}]\,
+\,g_{|n|}[
\cos{(\frac{|n|kz)}{2})}-\sin{(\frac{|n|kz)}{2})}]\,\}
\varphi_{|n|}(x) \nonumber \\
&&\label{a5} \\
&&\varphi_{|n|}(x)\,=\,
\alpha_{|n|}(x)-\alpha_{-|n|}(x)
~~~,~~~~
\tilde{\varphi}_{|n|}(x)\,=\,
\beta_{|n|}(x)+\beta_{-|n|}(x) \nonumber \\
&&f_{|n|}\,=\,\frac{1}{2}(b_{|n|}\,+\,a_{|n|})
~~~,~~~~g_{|n|}\,=\,\frac{1}{2}(b_{|n|}\,-\,a_{|n|})~~~,~~~~
a_{|n|}^2+b_{|n|}^2\,=\,1 \nonumber \\
&&z=y_1,y_2~~~,~~~~k=k_1,k_2 \nonumber 
\end{eqnarray}
where $a_{|n|}$, $b_{|n|}$, $f_{|n|}$, $g_{|n|}$ are some constants. The 
absolute value signs enclosing $n$ in (\ref{a5}) are employed only to 
emphasize that those $n$'s are positive integers.  
Even and odd $n$ correspond to periodic and anti-periodic 
boundary conditions \cite{conformal}, respectively. The proceeding from 
the second line of (\ref{a5}) to the third one follows from the fact that 
all Kaluza-Klein modes are the same except their masses, and the sine 
and cosine terms in (\ref{a5}) for the same $n$ result in the same mass so 
they correspond to the same physical field. An expansion similar to 
(\ref{a5}) is also true for the 7th direction, $y_3$ while in that case 
passing from the first line to the second line of 
the equation does not hold since there is no symmetry similar to 
(\ref{a3a}) or (\ref{a3b}) for the 7th direction. So both positive and 
negative values of $n$ should be included in (\ref{a5}) for the expansion 
corresponding to $z=y_3$. The 4-dimensional 
parts of the Kaluza-Klein modes are taken to transform, under (\ref{a3a}) 
and (\ref{a3b}), as 
\begin{eqnarray}
&&\varphi_{n,m,r}(x)\,\rightarrow\,\xi^{\lambda_n}{\cal 
CPT}\,\varphi_{n,m,r}(-x)~~~~
\mbox{as}~~~~x^a\,\rightarrow\,-x^a \label{aa6}
\\
&&\varphi_{n,m,r}(x)\,\rightarrow\,\xi^{\lambda_m}{\cal 
CPT}\,\varphi_{n,m,r}(-x)~~~~\mbox{as}~~~~x^b\,\rightarrow\,-x^b 
\label{ab6} \\
&&\varphi_{n,m,r}(x)\,\rightarrow\,\xi^{\lambda_n+\lambda_m}{\cal 
CPT}\,\varphi_{n,m,r}(x)~~~~
~\mbox{as}~~~~x^a\,\rightarrow\,-x^a 
~~,~~~~x^b\,\rightarrow\,-x^b 
\label{a6} \\
&&\lambda_n=\frac{i}{2}(-1)^{\frac{n}{2}}
~~~\lambda_m=\frac{i}{2}(-1)^{\frac{m}{2}}~~~~~a=0,1,2,3,5~~;~~b=0,1,2,3,6 
\nonumber 
\end{eqnarray}
where $n$, $m$, $r$ are the modes corresponding to $y_1$, $y_2$, 
$y_3$ directions, respectively; $\xi$ is some constant other than 1 
or -1, and ${\cal CPT}$ denotes the part of (4-dimensional) CPT 
transformation acting on the spinor part of the field. Here I take the 
extra dimensional reflections essentially act only on the positions of the 
fields while they do not act on the spinor parts of the fields. So it is 
more natural to take ${\cal CPT}$ rather than ${\cal PT}$ since 
${\cal CPT}\propto\gamma_5=i\gamma^0\gamma^1\gamma^2\gamma^3$ and commutes 
with extra dimensional gamma matrices while  
${\cal PT}\propto\gamma_5\gamma^2$ and does not commute with the extra 
dimensional gamma matrices \cite{Weinberg, clifford}. The natural choice 
for $\xi$ would be 1 or -1 if no restriction is imposed on the couplings 
of different Kaluza-Klein modes. On the contrary I want to impose coupling 
of specified modes with each other e.g. excluding diagonal coupling of 
the modes while imposing the coupling of $n=4k+1$ modes to 
$n=4k+3$ modes in (\ref{a7c}) below. This imposition naturally follows 
when the transformations (\ref{a3a}), (\ref{a3b}) are supplemented by 
(\ref{aa6}), (\ref{ab6}):
The volume element in (\ref{a2}) is odd under either 
of (\ref{a3a}) or (\ref{a3b}) while the volume element plus the 
integration boundaries is even under either of these transformations. This 
requires the 4-dimensional part of the kinetic term in ${\cal L}_f$ be 
even under either of the transformations. The 4-dimensional kinetic term 
of ${\cal L}_f$ is invariant under the 4-dimensional CPT transformation. 
This guarantees the 4-dimensional kinetic term in ${\cal L}_f$ be even 
under extra dimensional part of the transformations in (\ref{a3a}), 
(\ref{aa6}) 
(or (\ref{a3b}), (\ref{ab6})). This, in turn, requires $n=4k+1$ modes 
couple to $n=4k+3$ modes in (\ref{a7c}).
The significance of this type 
of imposition on the form of the Lagrangian will be evident when we 
consider the resulting field spectrum in the following paragraphs. So 
$\xi$ is taken to be an arbitrary constant other than 1 or -1. It is 
evident that the possible values of $\lambda_{n(m)}$ are 
$\frac{i}{2}$, $-\frac{i}{2}$, $\frac{1}{2}$, $-\frac{1}{2}$. Hence the 
terms of the form $\bar{\psi}_{n_1,n_2}\psi_{m_1,m_2}$ are invariant under 
(\ref{aa6}) and/or (\ref{ab6}) only for specific values of $n$ and $m$. 
This point as well will be used to construct correct action functionals in 
the following part of the paper. I also give the transformation rule of 
the fields under the simultaneous application of (\ref{aa6}) and 
(\ref{ab6}) as a reference for the evaluation of $S_{fk2}$ given in 
(\ref{a9})
\begin{eqnarray} 
&&\varphi_{n,m,r}(x)\,\rightarrow\,\xi^{\mp\,1}{\cal 
CPT}\,\varphi_{n,m,r}(-x)~~~~\mbox{if}~~~~
n=4l+1~,~m=4p+1\nonumber \\
&&\varphi_{n,m,r}(x)\,\rightarrow\,{\cal 
CPT}\,\varphi_{n,m,r}(-x)~~~~\mbox{if}~~~~
n=4l+1~,~m=4p+3~~~~
\mbox{or}~~~n=4l+3~,~m=4p+1 \nonumber \\
&&\varphi_{n,m,r}(x)\,\rightarrow\,\xi^{\pm\,1}{\cal 
CPT}\,\varphi_{n,m,r}(-x)~~~~\mbox{if}~~~~
n=4l+3~,~m=4p+3 \label{a6a}\\
&&l,p=0,1,2,.....
\nonumber 
\end{eqnarray}

I adopt anti-periodic boundary 
conditions \cite{conformal} for the 5th and 6th directions while I take 
periodic boundary conditions for the 7th direction. So $n$'s in (\ref{a5}) 
are odd integers for the 5th and 6th directions, $y_1$ and $y_2$ while 
they are even integers for the 7th direction, $y_3$. The reason for 
adopting anti-periodic boundary conditions for $y_1$ and $y_2$, and 
periodic conditions for $y_3$ will be discussed in the paragraph after
Eq.(\ref{a4aa}). I also introduce the following symmetry transformations
\begin{eqnarray}
&&k_{1}y_{1}\;\rightarrow\;
k_{1}y_{1}\,+\,\pi \label{a4a} \\
&&k_{2}y_{2}\;\rightarrow\;
k_{2}y_{2}\,+\,\pi
\label{a4b}
\end{eqnarray}
One observes that 
\begin{eqnarray} 
&&\mbox{as}~~~~ky\rightarrow\,ky+\pi \nonumber \\
&&i)~\mbox{if}~~n=4l+1~\Rightarrow 
~~(\,\cos{\frac{n}{2}ky}\,+\,\sin{\frac{n}{2}ky}\,)
\,\rightarrow\,(\,\cos{\frac{n}{2}ky}\,-\,\sin{\frac{n}{2}ky}\,) 
\nonumber \\
&&(\,\cos{\frac{n}{2}ky}\,-\,\sin{\frac{n}{2}ky}\,)
\,\rightarrow\,-\,(\,\cos{\frac{n}{2}ky}\,+\,\sin{\frac{n}{2}ky}\,) 
\nonumber \\
&&ii)~\mbox{if}~~n=4l+3~\Rightarrow 
~~(\,\cos{\frac{n}{2}ky}\,+\,\sin{\frac{n}{2}ky}\,)
\,\rightarrow\,-\,(\,\cos{\frac{n}{2}ky}\,-\,\sin{\frac{n}{2}ky}\,) 
\nonumber \\
&&(\,\cos{\frac{n}{2}ky}\,-\,\sin{\frac{n}{2}ky}\,)
\,\rightarrow\,\,(\,\cos{\frac{n}{2}ky}\,+\,\sin{\frac{n}{2}ky}\,) 
\label{a4aa}
\end{eqnarray}
An important observation at this point is that there is no nontrivial even 
or odd parity $\varphi$ under (\ref{a4a}) or (\ref{a4b}) 
if it obeys anti-periodic boundary conditions because in this case 
(\ref{a4a}) or (\ref{a4b}) effectively corresponds to a transformation of 
the form of $u\rightarrow\,\frac{\pi}{2}+u$ where $u=\frac{ky}{2}$, that 
is, the transformation induces a group of order four rather than a group 
of order two (that would be the case for $v\rightarrow\,\pi+v$, $v=ky$). 
In other words 
$\varphi^E=\varphi+\varphi^{P}+\varphi^{PP}+\varphi^{PPP}$ is even while 
$\varphi^O=\varphi+\varphi^{PP}-\varphi^{P}-\varphi^{PPP}$ is odd
, where  $\varphi^{P}=K:\varphi$, $\varphi^{PP}=K:\varphi^{P}$,
$\varphi^{PPP}=K:\varphi^{PP}$, $K$ stands for either of the 
transformations (\ref{a4a}) or (\ref{a4b}). However, after using the 
explicit form of $\varphi$ in (\ref{a5}), one notices that both even and 
odd eigenvectors of $K$, $\varphi^E$, $\varphi^O$, are identically zero. 
So one can not construct the Lagrangian ${\cal L}_k$, in particular the 
terms that are quadratic in $\varphi$ e.g. kinetic terms,  out of 
linear combinations of $\varphi^E$ and $\varphi^O$.  However one observes 
from (\ref{a4aa}) that the quadratic terms that are even or odd under $K$ 
may be written in the form, $\varphi\varphi\pm\varphi^P\varphi^P$ or
$\varphi\varphi^P\pm\varphi^P\varphi$. 
In fact this is the reason for adopting anti-periodic boundary conditions 
for $y_1$ and $y_2$. As we will see $S_{fk1}$ in (\ref{a7c}) should 
contain only off-diagonally coupled Kaluza-Klein modes and $S_{fk1}$ 
should vanish after integration over extra dimension to use these modes 
for the Pauli-Villars like regularization at the length scales smaller 
than the size of the extra dimensions. This, in turn, requires the 
$n_1$ and $m_1$ should be summed as in the 
$\cos{\frac{n_1+m_1}{2}k_1y_1}$ term in (\ref{a7c}). In other words it 
requires $\varphi$ and $\varphi^P$ be in the either of the combinations 
$\varphi\varphi+\varphi^P\varphi^P$ or $\varphi\varphi^P-\varphi^P\varphi$ 
i.e. in a combination of the form 
$i\bar{\varphi}_{n_1}\gamma^\mu\partial_\mu\varphi_{m_1}
(\cos{\frac{n_1}{2}k_1y_1}\cos{\frac{m_1}{2}k_1y_1}
-\sin{\frac{n_1}{2}k_1y_1}\sin{\frac{m_1}{2}k_1y_1})$ as done in 
(\ref{a7a}-\ref{a7c}). Such a combination can be enforced only if the 
$\varphi$ transforms as in (\ref{a4aa}) i.e. if the corresponding 
dimensions obey anti-periodic boundary conditions. A similar argument is 
true for the boundary conditions for $y_2$. The boundary conditions in the 
direction of $y_2$ should be anti-periodic as well in order to make 
$S_{fk2}$ in (\ref{a9}) to be non-vanishing after integration over extra 
dimensions. On the other hand the role of $y_3$ is only to enable the 
modification of the volume element on the brane $k_1y_1=k_3y_3$ to induce 
the usual fermions through $S_{fk2}$. Taking 
anti-periodic boundary conditions for $y_3$ is unnecessary and only 
causes complications such as a possible mass term (of order of the 
inverse size of the dimension $y_3$) for 
the lowest mode in the direction of $y_3$ (that to be identified by the 
usual fermions) on contrary to the phenomenology. These observations will 
be used to construct the actions $S_{fk1}$ and $S_{fk2}$ in the following 
paragraphs. Next I write down $\varphi^P$ explicitly for later reference, 
\begin{equation}
\varphi^P(x,z)=
\sum_{|n|=1}^{\infty}
\{\,\pm\,f_{|n|}[
\cos{(\frac{|n|kz}{2})}-\sin{(\frac{|n|kz)}{2})}]\,\mp\,g_{|n|}[
\cos{(\frac{|n|kz)}{2})}+\sin{(\frac{|n|kz)}{2})}]\,\}
\varphi_{|n|}(x)
\label{a5a} 
\end{equation}
where 
$+$ and $-$ in $\pm$ stands for $n=4p+1$ and $n=4p+3$, respectively 
while $-$ and $+$ in $\mp$ stands for $n=4p+1$ and $n=4p+3$, respectively. 
I also note the following relation for later reference,
\begin{eqnarray}
\partial_z\varphi(x,z)&=&
\begin{array}{c} 
\frac{k}{2}\,\sum_{|n|=1}^\infty|n|\,\varphi_n^P ~~~~\mbox{if}~~~~n=4p+1 
\\ 
-\frac{k}{2}\sum_{|n|=1}^\infty|n|\,\varphi_n^P ~~~~\mbox{if}~~~~n=4p+3 
\end{array} 
\label{a6aa} \\ 
\partial_z\varphi^P(x,z)&=&
\begin{array}{c} 
-\frac{k}{2}\sum_{|n|=1}^\infty|n|\,\varphi_n ~~~~\mbox{if}~~~~n=4p+1 \\ 
+\frac{k}{2}\sum_{|n|=1}^\infty|n|\,\varphi_n ~~~~\mbox{if}~~~~n=4p+3 
\end{array} 
\label{a6ab} \\ 
&&\varphi\,=\,\sum_{|n|=1}^\infty\varphi_n~~~,~~~~
\varphi^P\,=\,\sum_{|n|=1}^\infty\varphi_n^P \nonumber 
\end{eqnarray}
where the explicit forms of $\varphi_n$
and $\varphi_n^P$ are evident from (\ref{a5}) and (\ref{a5a}).

\section{The model}

Once the background of the model is studied we are ready to formulate 
the model now. The space employed in this scheme is the one given in 
(\ref{a1}). I particularize the analysis to fermionic fields, and replace 
$\varphi$ by $\chi$. In this paper the action in the bulk will be taken to 
be invariant under the separate applications of the transformations 
(\ref{a3a}) (and (\ref{aa6})) and (\ref{a4a}) while it is broken on the 
brane $y_1=y_3$ by a small amount. On the other hand the action on the 
brane will be taken to be invariant under the separate (and the 
simultaneous) applications of (\ref{a4a}) and (\ref{a4b}), and the 
simultaneous combined application (\ref{a3a}), (\ref{a3b}) (and 
(\ref{aa6}), (\ref{ab6})). These symmetries together with anti-periodic 
boundary conditions in the 5th, 6th directions and periodic boundary 
conditions in the 7th direction will lead to a model with an inbuilt 
Pauli-Villars regularization scheme as we will see in the following 
paragraphs. 
 
\subsection{The Spectrum at the Scales Larger than the Sizes of Extra 
Dimensions}

First we consider the 4-dimensional part of the 
kinetic term (except the spin connection term) of (\ref{a2}) for fermion 
fields. I require the action be invariant under (\ref{a3a}) and 
(\ref{a4a}). I consider the zero mode in the $y_3$ direction in the 
following and take the other modes be very heavy. In other words I assume 
only the zero mode of $y_3$ be relevant to the phenomenology at the 
present relatively low energies that can be produced in current 
or near future accelerators. Then the requirement of the action be 
invariant under (\ref{a3a}) ( and (\ref{a6a}) ) implies the Kaluza-Klein 
(KK) modes with $n=4p+1$ couples to KK modes with $n=4l+3$, $p,l=0,1,2,..$ 
in the Lagrangian terms that are quadratic in $\chi$ and $\chi^P$ (e.g. in 
the kinetic terms). Then in the light of the discussion after (\ref{a4aa}) 
and the symmetry (\ref{a3a}) ( and (\ref{aa6}) ) the requirement of 
invariance of the quadratic terms under (\ref{a4a}) requires the 
corresponding action be  
\begin{eqnarray}
S_{fk1} 
&=& \int \,
\;d^4x\,\,d^3y\,
\cos^3{k_2y_2}\,\cos{k_3y_3}\frac{1}{2}
[{\cal L}_{fk11}\,+\,
{\cal L}_{fk12}]
\;+\,H.C. \label{a7}\\
&&
{\cal L}_{fk11}\,=\,
\frac{i}{4}[(\bar{\chi}_{(1)}\gamma^{\mu}\,\partial_\mu\chi_{(3)} 
+\bar{\chi_{(1)}}^P\gamma^{\mu}\,\partial_\mu\chi_{(3)}^P) 
\,+\,y_1\rightarrow-y_1] \label{a7a} \\
&&
{\cal L}_{fk12}\,=\,
\frac{i}{4}[(\bar{\chi}\gamma^{\mu}\,\partial_\mu\chi^P
-\bar{\chi}^P\gamma^{\mu}\,\partial_\mu\chi)+(y_1\rightarrow-y_1)] 
\label{a7b}
\end{eqnarray}
After inserting the explicit forms of $\chi$ and $\chi^P$ (by using 
(\ref{a5}) and (\ref{a5a})) one finds 
\begin{eqnarray}
{\cal L}_{fk1}&=&
\frac{1}{2}[{\cal L}_{fk11}\,+\,
{\cal L}_{fk12}]\nonumber \\
&=&
\sum_{n_1,m_1=1}^\infty 
\,A_{n1,m1}^{(1,3)}
\,i\bar{\chi}_{n_1}(x,y)\gamma^\mu\partial_\mu\chi_{m_1}(x,y)
\,\cos{\frac{n_1+m_1}{2}k_1y_1}\,+\,H.C.\, \label{a7cc} \\
S_{fk1} 
&=& \int \,
\;d^4x\,\,d^2y\,
\cos^3{k_2y_2}\,\cos{k_3y_3}
\sum_{n_1,m_1=1}^\infty 
\,A_{n1,m1}^{(1,3)}
\,i\bar{\chi}_{n_1}(x,y)\gamma^\mu\partial_\mu\chi_{m_1}(x,y) \nonumber \\
&&\times\,\int\,dy_1\,\cos{\frac{n_1+m_1}{2}k_1y_1}\,+\,H.C.\;=\,0 
\label{a7c} \\
&&A_{n1,m1}^{(1,3)}\,=\,
(f_{n1}^*g_{m1}+g_{n1}^*f_{m1}+f_{n1}^*f_{m1}-g_{n1}^*g_{m1}) \nonumber 
\end{eqnarray}
where $y=y_2,y_3$ in general, and $y=y_2$ for the zero mode in the 
direction of $y_3$.
Here the upper index $*$ denotes complex conjugate, $H.C.$ stands for 
Hermitian conjugate, and $f_n$, $g_n$'s 
are those given in (\ref{a5}). The subscripts $(1)$, $(3)$ above refer to 
the modes with $n=4p+1$ and $n=4p+3$, respectively; and the superscript 
$(1,3)$ denotes that one of the subscripts $n_1$, $m_1$ is given by 
$4p_1+1$ while the other by $4s_1+3$, where $p,s=0,1,2,.....$. 
The $y_1\rightarrow-y_1$ terms in the above equations stand for the terms 
obtained from the preceding ones by replacing $y_1$'s in that term by 
$-y_1$ and insures the invariance of the Lagrangian ${\cal L}_{fk1}$ under 
(\ref{a3a}). The values of $n_1$, $m_1$ in (\ref{a7b},\ref{a7c}) are fixed 
by the requirement of invariance under 
(\ref{a4a}), (\ref{aa6}), and are given by
\begin{equation}  
n_1\,=\,4l_1 \,+\,1~,~~ m_1\,=\,4p_1\,+\,3~~~~\mbox{or vice versa}~~~~~~~~
l_1,p_1=0,1,2,....... 
\label{a7d}
\end{equation}
It is evident that the integration in (\ref{a7c}) results in zero 
because $\int_0^{L_1} 
\,\cos{\frac{n_1+m_1}{2}k_1y_1}\,dy_1=0$
since $n_1+m_1\neq\,0$.

On the hyper-surface $k_3y_3=k_1y_1$, I assume the symmetry (\ref{a3a}) 
(and (\ref{aa6}) ) is broken by a small amount while there is an unbroken 
symmetry under the separate (and simultaneous) applications of 
(\ref{a4a}), (\ref{a4b}), and under the simultaneous application of 
(\ref{a3a}) and (\ref{a3b}) (and (\ref{aa6}) and (\ref{ab6}) ). Then in 
addition to (\ref{a7}) there are additional terms given by
\begin{eqnarray}
&&S_{fk2}\,=\epsilon\,\int\;d^4x\,\,d^3y\,
\delta(k_3y_3-k_1y_1)
\cos^3{k_2y_2}\,\cos{k_3y_3} 
\frac{1}{2}[{\cal L}_{fk21}\,+\,
{\cal L}_{fk22}]
\;+\,H.C. \label{a9}\\
&&
{\cal L}_{fk21}\,=\,
\frac{i}{8}
[(\bar{\chi}_{(1,3)}\gamma^\mu\,\partial_\mu\chi_{(1,3)}+ 
\bar{\chi}_{(1,3)}^{P1,P2}\gamma^\mu\,\partial_\mu\chi_{(1,3)}^{P1,P2}-
\bar{\chi}_{(1,3)}^{P1}\gamma^\mu\,\partial_\mu\chi_{(1,3)}^{P1}-
\bar{\chi}_{(1,3)}^{P2}\gamma^\mu\,\partial_\mu\chi_{(1,3)}^{P2})\nonumber \\ 
&&+\,
(y_{1,2}\rightarrow-y_{1,2})] 
\label{a9a} \\
&&{\cal L}_{fk22}\,=\,
\frac{i}{8}[(\bar{\chi}_{(1,3)}\gamma^{\mu}\,\partial_\mu\chi_{(1,3)}^{P1}
+\bar{\chi}_{(1,3)}^{P1}\gamma^{\mu}\,\partial_\mu\chi_{(1,3)}
-\bar{\chi}_{(1,3)}^{P2}\gamma^{\mu}\,\partial_\mu\chi_{(1,3)}^{P1,P2}
-\bar{\chi}_{(1,3)}^{P1,P2}\gamma^{\mu}\,\partial_\mu\chi_{(1,3)}^{P2}
\nonumber \\
&&+\bar{\chi}_{(1,3)}\gamma^{\mu}\,\partial_\mu\chi_{(1,3)}^{P2} 
+\bar{\chi}_{(1,3)}^{P2}\gamma^{\mu}\,\partial_\mu\chi_{(1,3)}
-\bar{\chi}_{(1,3)}^{P1}\gamma^{\mu}\,\partial_\mu\chi_{(1,3)}^{P1,P2}
-\bar{\chi}_{(1,3)}^{P1,P2}\gamma^{\mu}\,\partial_\mu\chi_{(1,3)}^{P1}
+\bar{\chi}_{(1,3)}^{P1}\gamma^{\mu}\,\partial_\mu\chi_{(1,3)}^{P2}
\nonumber \\
&&+\bar{\chi}_{(1,3)}^{P2}\gamma^{\mu}\,\partial_\mu\chi_{(1,3)}^{P1} 
+\bar{\chi}_{(1,3)}\gamma^{\mu}\,\partial_\mu\chi_{(1,3)}^{P1,P2}
+\bar{\chi}_{(1,3)}^{P1,P2}\gamma^{\mu}\,\partial_\mu\chi_{(1,3)})
+(y_{1,2}\rightarrow-y_{1,2})] \label{a9b}
\end{eqnarray}
where $\epsilon\,<<\,1$ is some constant that accounts for the breaking of 
the symmetry (\ref{a3a}) by a small amount. The superscripts $P1$, $P2$ 
refer to the $\chi$'s transformed under (\ref{a4a}), (\ref{a4b}), 
respectively. The subscripts $(1,3)$  refer to 
$n_1,m_1=4p_1+1$ and $n_2,m_2=4p_2+3$, $p_1,p_2=0,1,2,...$. 
After replacing the fields $\chi$, $\chi^P$ one finds
\begin{eqnarray}
S_{fk2}
&=&
\frac{\epsilon L_3}{4\pi}
\sum_{n_1,m_1=1}^\infty 
\sum_{n_2,m_2=1}^\infty 
A_{n1,m1}^{(1,1)}
A_{n2,m2}^{(3,3)}
\int d^4x
\,i\bar{\chi}_{n_1,n_2}\gamma^\mu\partial_\mu\chi_{m_1,m_2} 
\nonumber \\
&&
\times\,\int\,dy_1
[\,\cos{(\frac{n_1+m_1}{2}-1)k_1y_1}\,+
\,\cos{(\frac{n_1+m_1}{2}+1)k_1y_1}\,] \nonumber \\
&&\times\,\int\,dy_2
[\,3\cos{(\frac{n_2+m_2}{2}-1)k_2y_2}\,+
\,3\cos{(\frac{n_2+m_2}{2}+1)k_2y_2} \nonumber \\ 
&&+\,\cos{(\frac{n_2+m_2}{2}-3)k_2y_2}\,
+\,\cos{(\frac{n_2+m_2}{2}+3)k_2y_2}\,]
\label{a9c} \\
&&A_{n1,m1}^{(1,1)}\,=\,
(f_{n1}^*g_{m1}+g_{n1}^*f_{m1}+f_{n1}^*f_{m1}-g_{n1}^*g_{m1}) \nonumber \\
&&\tilde{A}_{n2,m2}^{(3,3)}\,=\,
(f_{n2}^*g_{m2}+g_{n2}^*f_{m2}
+f_{n2}^*f_{m2}-g_{n2}^*g_{m2}) \nonumber 
\end{eqnarray}
The superscript $(1,1)$ in (\ref{a9c}) refers to 
the fact that the modes with $n_1=4p_1+1$ couple to those 
with $m_1=4l_1+1$ while the superscript $(3,3)$ refers to that 
the modes with $n_2=4p_2+3$ couple to the modes with
$m_2=4l_2+3$. In other words the values of $n_1$, 
$m_1$, $n_2$, $m_2$ are fixed by the requirement of invariance under 
(\ref{a4a}), (\ref{a4b}), (\ref{a6a}), and are given by
\begin{eqnarray}  
&&n_1\,=\,4p_1 \,+\,1~,~~ m_1\,=\,4l_1\,+\,1
~~,~~n_2\,=\,4p_2 \,+\,3~,~~ m_2\,=\,4l_2\,+\,3 \nonumber \\
&&~\mbox{or}~~~~
n_1\,=\,4p_1 \,+\,3~,~~ m_1\,=\,4l_1\,+\,3
~~,~~n_2\,=\,4p_2 \,+\,1~,~~ m_2\,=\,4l_2\,+\,1 
\label{a10a} \\
&&l_1,p_1=0,1,2,.......  \nonumber 
\end{eqnarray}
Due to the periodicity of the cosine functions (\ref{a9c}) is 
non-zero after integration over extra dimensions only when the argument of 
the cosines in (\ref{a9c}) are zero, that is, when 
\begin{eqnarray}
&&n_1+m_1-2=(4l_1+1)+(4p_1+1)-2=0~~\Rightarrow~~l_1=p_1=0 
~\Rightarrow~~n_1=m_2=1 
\nonumber \\
&&n_2+m_2-6=(4l_2+3)+(4p_2+3)-6=0
~\Rightarrow~l_2=p_2=0 
~\Rightarrow~~n_2=m_2=3 
\label{a10b}
\end{eqnarray}
So the integral in (\ref{a9c}) gives
\begin{equation}
S_{fk2}\,=\,
\frac{\epsilon L_1L_2L_3}{4\pi}
(f_{1}^*g_{1}+g_{1}^*f_{1}+f_{1}^*f_{1}-g_{1}^*g_{1})
(f_{3}^{\prime*}g_{3}^\prime+g_{3}^{\prime*}f_{3}^\prime+f_{3}^{\prime*}
f_{3}^\prime-g_{3}^{\prime*}g_{3}^\prime)
\int \,d^4x
\,i\bar{\chi}_{13}\gamma^\mu\partial_\mu\chi_{13}
\label{a9} 
\end{equation}
where the primes on $f_3^\prime$, $g_3^\prime$ are introduced to point out 
that these are the Fourier expansion coefficients in $y_2$ direction while 
$f_1$, $g_1$ here are the Fourier expansion coefficients in $y_1$ 
direction.  

In other words at energies smaller than 
$\sim\,\frac{1}{L_{1(2)}}$ only 
$\chi_{13}$ is observed. Further if $n_3=0$ is identified by the usual 
particles (and the other modes in the $y_3$ direction are assumed to be 
very heavy) then $\chi_{130}$ (where $n_3=0$ is the zero mode 
corresponding to $y_3$ direction) is the 
only particle observed at present energies and it is identified by a
usual (standard model) fermion. Note that the higher modes 
$\chi_{n_1,n_2,0}$ will not be observed at length scales larger then 
$L_{1(2)}$ even when they are somehow produced on contrary to the usual 
way of getting rid of higher Kaluza-Klein modes by taking them very 
massive (compared to the energy scales attainable at current experiments). 
Moreover the matter action (hence the Lagrangian) is multiplied by the 
small parameter $\epsilon$ at scales larger than the size of the extra 
dimensions. This may explain why gravitational force is so smaller than 
the other forces since the Lagrangian enters the Einstein equations 
through energy-momentum tensor. Another point worth to mention is that the 
dimension of the space employed here (i.e 7) is the minimum dimension that 
this scheme can be applied as is evident from the argument given in the 
preceding paragraphs. Fifth dimension, $y_1$ is necessary to make 
the contribution due to $S_{fk1}$ (that is used for regularization at 
smaller length scales) be vanishing at relatively large length scales by 
the requirement of the invariance of the action under (\ref{a3a}) and 
(\ref{a4a}). The sixth dimension, $y_2$ is necessary to induce the 
non-zero $i\bar{\chi}_{13}\gamma^\mu\partial_\mu\chi_{13}$ in (\ref{a9}) 
by the requirement of the invariance of the action under the separate (and 
simultaneous) applications of (\ref{a4a}), (\ref{a4b}), and the 
simultaneous application of (\ref{a3a}) and (\ref{a3b}) ( and (\ref{aa6} 
and (\ref{ab6})). The role of the seventh dimension, $y_3$ is to change 
the factor $\cos{k_3y_3}$ in the volume element to $\cos{k_1y_1}$ by the 
delta function so that the non-zero contribution to $S_{fk2}$ at large 
length scales through the diagonal term 
$i\bar{\chi}_{13}\gamma^\mu\partial_\mu\chi_{13}$ may be 
induced. In fact this also explains why the transformations 
(\ref{a3a}), (\ref{a3b}), (\ref{a4a}), (\ref{a4b}) do not act on $y_3$. 
The only role of $y_3$ is to change the form of the volume element so that 
the diagonal term $i\bar{\chi}_{13}\gamma^\mu\partial_\mu\chi_{13}$ in 
(\ref{a9}) is induced. A non-trivial transformation of $y_3$ under these 
transformation would only make the model more complicated.

Next consider a possible mass for $\chi_{130}(x)$ that may be induced by 
the extra dimensional pieces of the kinetic terms in 
$S_{fk1}$  and $S_{fk2}$. First consider the spin connection terms of the 
form $\bar{\chi}\Gamma^A\,\omega_A\chi$ where $\omega_A=
(e^a_B \partial_A e^{Bb} +
e^a_B e^{Cb}\Gamma_{CA}^B)[\gamma^a,\gamma^b]$, 
($A,B,C,a,b=0,1,2,3,5,6,7$). 
Although the spin connection 
terms do not have the form of a mass term in the simplest scheme 
where $f_n$, $g_n$ are simple real numbers one may obtain mass terms if 
we allow a more general form for 
$f_n$, $g_n$, which is compatible with the 4-dimensional local Lorentz 
invariance, that is,
\begin{equation}
f_n=f_{0n}+
\sum\Gamma^af_{a\,n}
+\sum\Gamma^a\Gamma^bf_{ab\,n}
+\sum\Gamma^a\Gamma^b\Gamma^cf_{abc\,n}
~,~~~\{\Gamma^a,\Gamma^b\}=2g^{ab}~,~~~
a,b,c=5,6,7
\label{a10b}
\end{equation}
where a similar expression may be written for $g_n$ as well.
The non-zero values of the 
spin connection $\omega_A$ are 
$\omega_5\propto\frac{\tan{k_2y_2}}{\cos{k_3y_3}}$, 
$\omega_6\propto\sin{k_3y_3}$, 
$\omega_7\propto \cos^2{k_3y_3}\sin{k_3y_3}$. After integrating over 
$y_1$, $y_2$, $y_3$ they give zero for both of the terms of the form 
$S_{fk1}$ and $S_{fk2}$ because the symmetries (\ref{aa6}, \ref{ab6}) set 
the overall extra dimensional contribution from $\bar{\chi}\Gamma^A\chi$ 
to be a cosine while the spin connection terms contain one sine term so 
that the overall extra dimensional contribution is a sine that gives zero 
after integration over extra dimensions. So spin connection 
terms are already are not relevant for the 4-dimensional mass terms. 
Next consider a possible mass term that may be induced by the extra 
dimensional derivative terms of the form
$\bar{\chi}\Gamma^a\,\partial_a\chi$ where $a=5,6,7$ ($x_5=y_1$, 
$x_6=y_2$, $x_7=y_3$) in the kinetic terms. It is evident from 
(\ref{a6aa}) and (\ref{a6ab}) that the extra dimensional 
pieces of the kinetic terms (due to derivatives) in the Lagrangian does 
not obey the symmetry under the simultaneous application of (\ref{a3a}) and 
(\ref{a3b}). So they are not allowed. In fact explicit evaluation of these 
terms give zero identically due to the same reason as the null 
contribution of the spin connection term to mass. So no masses are induced 
due to the extra dimensional piece of the kinetic terms. However one may 
introduce a mass through a term $m\bar{\chi}\chi$ (or a fermion-Higgs 
interaction term $m\bar{\chi}\phi\chi$ ) on the brane $k_1y_1=k_3y_3$ in 
the same as done for getting the 
$\bar{\chi}_{13}\gamma^\mu\partial_\mu\chi_{13}$ in (\ref{a9}). 

\subsection{General Considerations on the Spectrum}

Before discussing the field spectrum at the scales smaller than the sizes 
of extra dimensions I write the fields and the Lagrangian 
in a simpler form so that the discussion of the following parts 
becomes simpler. The results obtained in this subsection will especially 
be important for the discussion of the field spectrum at the scales 
smaller than the sizes of extra dimensions while the results hold for 
all scales. However the results obtained here are not crucial for 
the discussion of the preceding subsection. Moreover the results of the 
preceding subsection will be used to clarify the general statements 
given here. So this is the right point to discuss the results obtained 
here. 

In the general simple Kaluza-Klein (KK) prescription all 
Kaluza-Klein modes correspond to  distinct elementary particles that are 
independent of each other (except sharing the same internal properties). 
On the other hand the KK modes in this scheme in general mix with each 
other through off-diagonal couplings in the kinetic terms. It is 
evident from the discussion given in the preceding subsection that the 
modes in this scheme belong to four different sets. The sets (for $n_3=0$) 
are; i) $n_1=4p_1+1$, $n_2=4p_2+1$, ii) $n_1=4p_1+1$, $n_2=4p_2+3$,
iii) $n_1=4p_1+3$, $n_2=4p_2+1$, i) $n_1=4p_1+3$, $n_2=4p_2+3$. The sets 
that are relevant for us are ii), iii), iii) because we are concerned with 
the modes that couple to $\psi_{130}$ (i.e. to the usual standard model 
fermion in this construction). The modes in each set can not be 
identified with distinct physical fields because all modes in the set are 
entangled. So each set must be identified with a particular field (or 
particle). At this point I make two plausible 
assumptions to simplify the analysis. I take the extra dimensions be 
related to the internal properties of fields. I  also take the internal 
properties of the fields be independent of their 4-dimensional coordinates.
These two conditions greatly simplify the Fourier decomposition of a field 
corresponding to one of these sets. To be specific, for example, consider 
the modes with $n_1=4p_1+1$. The corresponding Fourier decomposition may 
be written as
\begin{eqnarray}
\chi_{(1)}\,
(x,y)\,=\,
\sum_{p=0}^\infty 
[\,
a_{4p+1}^{(1)}(y_2,y_3)
\cos{\frac{4p+1}{2}k_1y_1}\,+\,
b_{4p+1}^{(1)}(y_2,y_3)
\sin{\frac{4p+1}{2}k_1y_1}\,]
\chi_n^{(1)}(x^\mu)
\label{a14aa} 
\end{eqnarray}
In order to be able to 
satisfy the condition that a field at different 4-dimensional coordinates
has the same internal properties  
the composition in (\ref{a14aa}) should reduce to the 
following form
\begin{eqnarray}
\chi_{(1)}\,
(x,y)
&=&\chi_1(x)\,F(y) \label{a16aa}\\
F(y)&=&\sum_{p=0}^\infty 
[\,
a_{4p+1}^{(1)}(y_2,y_3)
\cos{\frac{4p+1}{2}k_1y_1}\,+\,
b_{4p+1}^{(1)}(y_2,y_3)
\sin{\frac{4p+1}{2}k_1y_1}\,]
\label{a16aaa} 
\end{eqnarray}
provided that extra dimensions are 
identified with internal properties 
of particles. The consideration of this argument in a more concrete form 
through the study of $S_{fk1}$, $S_{fk2}$ may be more instructive. So I 
give such an analysis below.

First consider $S_{fk1}$. The general form of 
$S_{fk1}$ is
\begin{equation}
S_{fk1} \,\sim\,\int \,\cos^3{k_2y_2}\,\cos{k_3y_3}
\,[i\bar{\chi}_{(1)}\gamma^{\mu}\,\partial_\mu
\chi_{(3)} + 1\,\leftrightarrow \,3] \;d^4x\,\,d^3y \label{a13}
\end{equation}
where the terms with the upper indices $P$ in (\ref{a7a},\ref{a7b}) are 
skipped because the complete form of $S_{fk1}$ is not necessary for the 
analysis given here. The simple form given in (\ref{a13}) is enough to see 
the essential points in this subsection. Here
\begin{eqnarray}
\chi_{(1)}\,
(x,y)
&=&
\sum_{p=0}^\infty 
[\,
a_{4p+1}^{(1)}(y_2,y_3)
\cos{\frac{4p+1}{2}k_1y_1}\,+\,
b_{4p+1}^{(1)}(y_2,y_3)
\sin{\frac{4p+1}{2}k_1y_1}\,]
\chi_n^{(1)}(x^\mu)
\label{a14a} \\
\chi_{(3)}(x,y) 
&=& 
\sum_{p=0}^\infty 
[\,a_{4p+3}^{(3)}(y_2,y_3)\cos{\frac{4p+3}{2}k_1y_1}\,+\,
b_{4p+3}^{(3)}(y_2,y_3)\sin{\frac{4p+3}{2}k_1y_1}\,]
\chi_n^{(3)}(x^\mu)
\label{a14b}
\end{eqnarray}
After considering (\ref{a13}) and (\ref{a14a},\ref{a14b})
one observes that all modes in $\chi_{(1)}$ are mixed with each other and 
the same is true for $\chi_{(3)}$. So it is impossible to entangle the 
modes in these states as different particles. In other words one should 
treat $\chi_{(1)}$ or $\chi_{(3)}$ as a single entity. Extra dimensions 
are related to the internal properties of the elementary particles. 
Elementary particles at different 4-dimensional coordinates do not have 
different internal properties (at least at the scales reached by 
current experiments). If we assume this to hold at all scales in the 
4-dimensional coordinates it implies that 
\begin{equation}
\chi_n^{(1)}\,=\,  
\chi_m^{(1)}
~~~~,~~~~  
\chi_n^{(3)}\,=\,  
\chi_m^{(3)} ~~~~~~~~\mbox{for all}~n,m
\label{a15}
\end{equation}
Then (\ref{a14a}) and (\ref{a14b}) become
\begin{eqnarray}
\chi_{(1)}\,
(x,y)
&=&\chi_1(x)\,F(y) \label{a16a}\\
F(y)&=&\sum_{p=0}^\infty 
[\,
a_{4p+1}^{(1)}(y_2,y_3)
\cos{\frac{4p+1}{2}k_1y_1}\,+\,
b_{4p+1}^{(1)}(y_2,y_3)
\sin{\frac{4p+1}{2}k_1y_1}\,]
\label{a16aa} \\
\chi_{(3)}(x,y) 
&=& \chi_3(x)\,G(y) \label{a16b} \\
G(y)&=&\sum_{p=0}^\infty 
[\,
a_{4p+3}^{(3)}(y_2,y_3)
\cos{\frac{4p+3}{2}k_1y_1}\,+\,
b_{4p+3}^{(3)}(y_2,y_3)
\sin{\frac{4p+3}{2}k_1y_1}\,]
\label{a16ba}
\end{eqnarray}
In a similar way the general form of $S_{fk2}$ is
\begin{equation}
S_{fk2} \,\sim\,\frac{\epsilon}{k_3}\int \,\cos^3{k_2y_2}\,\cos{k_1y_1}
\delta(x_3-x_1)\,[
i\bar{\tilde{\chi}}_{(1,3)}\gamma^{\mu}\,\partial_\mu\tilde{\chi}_{(1,3)}
+i\bar{\tilde{\chi}}_{(3,1)}\gamma^{\mu}\,\partial_\mu\tilde{\chi}_{(3,1)}
] \;d^4x\,\,d^3y \label{a17}
\end{equation}
Here
\begin{eqnarray}
\tilde{\chi}_{(r,q)}\,
(x,y)
&=&\chi_{(r,q)}(x)\,\tilde{F}_r(y_1,y_3)
\tilde{G}_q(y_2,y_3) 
\label{a18a} \\
\tilde{F}_r(y_1,y_3)&=&\sum_{p=0}^\infty 
[\,
\tilde{a}_{4p+r}^{(1)}(y_3)
\cos{\frac{4p+r}{2}k_1y_1}\,+\,
\tilde{b}_{4p+r}^{(1)}(y_3)
\sin{\frac{4p+r}{2}k_1y_1}\,]
\label{a18b} \\
\tilde{G}_q(y_2,y_3)&=&\sum_{p=0}^\infty 
[\,
\tilde{a}_{4p+q}^{(3)}(y_3)
\cos{\frac{4p+q}{2}k_2y_2}\,+\,
\tilde{b}_{4p+q}^{(3)}(y_3)
\sin{\frac{4p+q}{2}k_2y_2}\,]
\label{a18c} \\
&&r,q=1,3~,~~r\neq q  \nonumber
\end{eqnarray}
where $a_n$'s, $b_n$'s in (\ref{a16aa},\ref{a16ba}) are changed into 
$\tilde{a}_n$'s, $\tilde{b}_n$'s since the fields in this case are 
confined into a subspace of the whole space, and all modes in the 
direction of $y_2$ contribute to $\chi_{(1)}$ in $S_{fk1}$ while only the 
modes with $n_2=4p_2+1$ contribute to $S_{fk2}$ when $n_1=4p_1+3$ and the 
modes $n_2=4p_2+3$ contribute when $n_1=4p_1+1$. 

\subsection{The spectrum at the scales smaller than the sizes of extra 
dimensions}

After the study of the technical points given above we return to the main 
discussion. Now see what happens when one goes to the scales smaller than 
the size of the extra dimensions $L_{1(2)}$. In scales smaller than the 
size of $L_{1(2)}$ all Kaluza-Klein modes in the corresponding 
direction $y_{1(2)}$ are observed since conformal 
factors $\cos{k_1y_1}$ and $\cos{k_2y_2}$ can not hide these modes any 
more. However it has been shown in the preceding section that these modes 
for $n_3=0$ reduce to 
$\chi_{130}(x)$, $\chi_{110}(x)$, $\chi_{310}(x)$, $\chi_{330}(x)$ from a 
4-dimensional point of view under the assumptions 
that extra dimensions can be identified by internal properties of fields 
and internal properties of particles are the same at all 4-dimensional 
coordinates (once their extra dimensional coordinates are kept fixed).
In the light of above analysis let us consider  the kinetic term of 
$\chi_{130}$ in the scales smaller than the sizes of 
the extra dimensions. 

\subsubsection{Outside the brane $k_1y_1=k_3y_3$}

At the points 
$k_1y_1\neq\,k_3y_3$ the only contribution 
is due to (\ref{a7cc}) where in the light of the preceding section the 
Lagrangian may be written as 
\begin{eqnarray}
{\cal L}_{fk1} &=&
\sum_r i\bar{\chi}_{130}(x)\gamma^{\mu}\,\partial_\mu\chi_{3r0}(x) 
\,\sum_{p,s=0}^\infty \tilde{A}_{ps}^{(r)}(y_2)
\cos{2(p+s+1)k_1y_1^\prime}
\label{a19} \\
\tilde{A}_{ps}^{(r)}(y_2)&=&
f_{4p+3}^{(3)*}(y_2)g_{4s+r}^{(r)}(y_2)+g_{4p+3}^{(3)*}(y_2)
f_{4s+r}^{(r)}(y_2)
+f_{4p+3}^{(3)*}(y_2)f_{4s+r}^{(r)}(y_2)
-g_{4p+3}^{(3)*}(y_2)g_{4s+r}^{(r)}(y_2) \nonumber
\end{eqnarray}
where, on contrary to (\ref{a7a}-\ref{a7c}), the $y_2$ dependence is 
expressed explicitly and $r=1,3$ stands for the modes 
$n_2=4p_2+1$, $n_2=4p_2+3$, 
respectively; and the upper indices $(1,3)$ in (\ref{a7cc}) are 
suppressed here because the fact that the modes $n_1=4p_1+1$ couple to 
the modes $m_1=4s_1+3$ is evident in (\ref{a19}). Eq.(\ref{a19}) 
may be written in the form 
\begin{eqnarray} {\cal L}_{fk1} &=&
\frac{i}{2}\lim_{x^\prime\rightarrow x}
\partial_\mu\;\left(
\,\bar{\chi}_{130}(x^\prime),
\bar{\chi}_{310}(x^\prime)
\bar{\chi}_{330}(x^\prime)
\,\right)\,{\bf M}\,
\gamma^{\mu}
\left( 
\begin{array}{c}
\chi_{130}(x)\\
\chi_{310}(x) \\
\chi_{330}(x)
\end{array}
\right)
\label{a19b} 
\end{eqnarray}
Here
\begin{eqnarray}
{\bf M}&=&\left(
\begin{array}{ccc}
0&{\cal B}&{\cal C} \\
{\cal B}&0&0 \\
{\cal C}&0&0
\end{array}
\right) \label{a19ba} \\
&&{\cal B}=\sum_{p,s=0}^\infty \tilde{A}_{ps}^{(1)}
\cos{2(p+s+1)k_1y_1^\prime} \label{a19bb}~~,~~~
{\cal C}=\sum_{p,s=0}^\infty \tilde{A}_{ps}^{(3)}
\cos{2(p+s+1)k_1y_1^\prime} \label{a19bc} 
\nonumber
\end{eqnarray}
The diagonalization of ${\bf M}$ in (\ref{a19ba}) results in 
\begin{eqnarray}
{\cal L}_{fk1} \,=\,i
B(y)[\bar{\psi}_1(x)\gamma^\mu\,\partial_\mu\psi_1(x)
\,-\,\bar{\psi}_2(x)\gamma^\mu\,\partial_\mu\psi_2(x)\,] \label{a20} 
\end{eqnarray}
\begin{eqnarray}
\psi_1&=&\frac{1}{\sqrt{2}}
[\chi_{130}+
(\cos{\theta}\chi_{310}+\sin{\theta}\chi_{330})] \label{a20a} \\
\psi_2&=&\frac{1}{\sqrt{2}}
[\chi_{130}-
(\cos{\theta}\chi_{310}+\sin{\theta}\chi_{330})]
\label{a20a} \\
&&\cot{\theta}=\frac{{\cal B}}{{\cal C}}~~~,~~~
B(y)=\frac{1}{4}\sqrt{({\cal B}^2+{\cal C}^2)}~~,~~~y=y_1,y_2 \label{a20b}
\end{eqnarray}
There is another state $\psi_3\,=\,
\sin{\theta}\chi_{310}-\cos{\theta}\chi_{330}$ but this does not 
contribute to (\ref{a20}). So it is an auxiliary field. Although the sign 
of the kinetic term of $\psi_2$ in (\ref{a20}) is opposite of a usual 
fermion (and so it is a ghost-like field) it does not suffer from the 
problems of the usual ghosts. $\psi_1$ or $\psi_2$ in (\ref{a20}) can not 
be introduced or removed from (\ref{a20}) separately because (\ref{a20}) 
follows from the couplings of $\chi_{130}$, $\chi_{310}$, $\chi_{330}$. So 
$\psi_1$, $\psi_2$ form a single system. For example in this case 
$\psi_1$, $\psi_2$ may be considered as the components of a single field 
with a 8-component spinor and the gamma matrices given by 
$\gamma^\mu\odot\tau_3$ where $\odot$ 
denotes tensor product and $\tau_3$ is the third Pauli matrix. This solves 
the problem of negative norm for $\psi_2$ because there is single norm 
i.e. that of the system composed of $\psi_1$, $\psi_2$. Moreover 
since $\psi_1$ and $\psi_2$ have the same internal space properties and 
they form a single system they may be assigned the same 4-momentum with 
positive energy, and this solves the negative energy problem of 
$\psi_2$. However the extension of this argument to the fields other than 
the fermions is not straightforward and requires additional study.

\subsubsection{On the brane $k_1y_1=k_3y_3$}

On the brane $k_1y_1=k_3y_3$ one should include the effect of 
${\cal L}_{fk2}$ as well. To see the situation near $k_1y_1=k_3y_3$ I 
consider a tiny patch on the brane and integrate ${\cal L}_{fk1}$,
${\cal L}_{fk2}$ over that patch while the dependence on $x^\mu$ and 
$y_2$ are point-wise. Then I form a mixing matrix similar to (\ref{a19ba}) 
to the study the resulting spectrum. For convenience I change the 
parameters $y_1$, $y_3$ to $u=k_1y_1-k_3y_3$, 
$v=k_1y_1+k_3y_3$. I take the patch to be a rectangular area
on the line $k_1y_1=k_3y_3$ and on its neighborhood, of the width 
and length $2\Delta$ and $\Delta^\prime$, given by
\begin{equation} 
-\Delta\leq\,u\,\leq\Delta~,~~~ v\leq\,v^\prime\,\leq\,v+\Delta^\prime~~~,
~~~~u=k_1y_1-k_3y_3~,~~v=k_1y_1+k_3y_3 \label{b1}
\end{equation} 
From a 4-dimensional perspective the effective Lagrangian at the scales 
smaller than the size of the extra dimensions may be taken as the original 
Lagrangian times the extra dimensional conformal factors in the volume 
element. The 
inclusion of these terms in the volume element is essential because the 
$\cos{k_3y_3}$ term in the volume element is effected by the integration 
over $y_3$ due to the $\delta$ function in ${\cal L}_{fk2}$. So I consider 
the extra dimensional terms in the volume element times the Lagrangian in 
the following.
The corresponding term for ${\cal L}_{fk1}$ is
\begin{eqnarray} 
&&i\sum_{r}\bar{\chi}_{130}(x)\gamma^{\mu}\,\partial_\mu\chi_{3r0}(x) 
\,\cos^3{k_2y_2}
\sum_{p,s=0}^\infty 
A_{ps}^{(r)}(y_2) \nonumber \\
&&\times\,
\int_v^{v+\Delta^\prime}dv^\prime
\int_{-\Delta}^{\Delta}\,du
\cos{\frac{1}{2}[2(p+s)+1](v^\prime+u)}
\,\cos{\frac{1}{2}(v^\prime-u)} \nonumber \\
&=&\frac{1}{2}\sum_r 
i\bar{\chi}_{130}(x)\gamma^{\mu}\,\partial_\mu\chi_{3r0}(x) 
\,\cos^3{k_2y_2}
A_{ps}^{(r)}(y_2) \nonumber \\
&&\times\,\int_v^{v+\Delta^\prime}dv^\prime\int_{-\Delta}^{\Delta}\,du
\{\cos{[(p+s)u+(p+s+1)v^\prime]}
+\cos{[(p+s+1)u+(p+s)v^\prime]} \} \nonumber \\
&=&\sum_r 
i\bar{\chi}_{130}(x)\gamma^{\mu}\,\partial_\mu\chi_{3r0}(x) 
\,\cos^3{k_2y_2} A_{ps}^{(r)}(y_2) \nonumber \\
&&\times\,\frac{1}{(p+s)(p+s+1)}[\,
\sin{(p+s)\Delta}\,\sin{(p+s+1)v^\prime}\mid_v^{v+\Delta^\prime}
+\sin{(p+s+1)\Delta}\,\sin{(p+s)v^\prime}\mid_v^{v+\Delta^\prime}] 
\nonumber \\
&\simeq&\sum_r 
i\bar{\chi}_{130}(x)\gamma^{\mu}\,\partial_\mu\chi_{3r0}(x) 
\,\cos^3{k_2y_2} A_{ps}^{(r)}(y_2) \nonumber \\
&&\times\,\frac{\Delta^\prime}{(p+s)(p+s+1)}[\,
\sin{(p+s)\Delta}\,\cos{(p+s+1)(k_1y_1+k_3y_3)} \nonumber \\
&&+\sin{(p+s+1)\Delta}\,\cos{(p+s)(k_1y_1+k_3y_3)}]
\label{ap1} 
\end{eqnarray}
Here
\begin{eqnarray}
&&
A_{ps}^{(r)}\,=\,
\sum_{p_2=0}^{\infty}
\{\,f^\prime_{p,|4p_2+3|}[
\cos{(\frac{|4p_2+3|k_2y_2}{2})}+\sin{(\frac{|4p_2+3|k_2y_2)}{2})}]\nonumber 
\\
&&+\,g^\prime_{p,|4p_2+3|}[
\cos{(\frac{|4p_2+3|k_2y_2)}{2})}-\sin{(\frac{|4p_2+3|k_2z_2)}{2})}]\,\}^* 
\nonumber\\
&&\times\,\sum_{s_2=0}^{\infty}
\{\,f^\prime_{s,|4s_2+r|}[
\cos{(\frac{|4s_2+r|k_2y_2}{2})}+\sin{(\frac{|4s_2+r|k_2y_2)}{2})}]\nonumber 
\\
&&+\,g^\prime_{s,|4s_2+r|}[
\cos{(\frac{|4s_2+r|k_2y_2)}{2})}-\sin{(\frac{|4s_2+r|k_2z_2)}{2})}]\,\}
\label{ap1a} 
\end{eqnarray}
where $r=1$ or $r=3$ stands for the modes in the direction of $y_2$ with 
$n_2=4l+1$ or $n_2=4l+3$, $l=0,1,2,...$; respectively; and the primes 
over $f_{n}$'s, $g_n$'s do not have the same meaning as those in 
(\ref{a9}) and stand for the fact that they are in general not 
$f_n$'s, $g_n$'s themselves, rather their linear combinations. The 
corresponding term for ${\cal L}_{fk2}$ is 
\begin{eqnarray}
&&i\epsilon\,\bar{\chi}_{130}(x)\gamma^{\mu}\,\partial_\mu\chi_{130}(x) 
\,\cos^3{k_2y_2}
\sum_{p_1,s_1=0}^\infty
\tilde{A}_{p_1s_1}^{(1,1)} \,
\sum_{p_2,s_2=0}^\infty
\tilde{A}_{p_2s_2}^{(3,3)}
\nonumber \\
&&\times\,\cos{[2(p_2+s_2)+3]k_2y_2}\,
\int_v^{v+\Delta^\prime}dv^\prime
\int_{-\Delta}^{\Delta}\,du
\cos{\frac{1}{2}[2(p_1+s_1)+1](v^\prime+u)}\,\delta(u)
\,\cos{\frac{1}{2}(v^\prime-u)} \nonumber \\
&+&i\epsilon\,\bar{\chi}_{310}(x)\gamma^{\mu}\,\partial_\mu\chi_{310}(x) 
\,\cos^3{k_2y_2}
\sum_{p_1,s_1=0}^\infty
\tilde{A}_{p_1s_1}^{(3,3)} \,
\sum_{p_2,s_2=0}^\infty
\tilde{A}_{p_2s_2}^{(1,1)}
\nonumber \\
&&\times\,\cos{[2(p_2+s_2)+1]k_2y_2}\,
\int_v^{v+\Delta^\prime}dv^\prime
\int_{-\Delta}^{\Delta}\,du
\cos{\frac{1}{2}[2(p_1+s_1)+3](v^\prime+u)}\,\delta(u)
\,\cos{\frac{1}{2}(v^\prime-u)} \nonumber \\
&=&i\epsilon\,\bar{\chi}_{130}(x)\gamma^{\mu}\,\partial_\mu\chi_{130}(x) 
\,\cos^3{k_2y_2}
\sum_{p_1,s_1=0}^\infty
\tilde{A}_{p_1s_1}^{(1,1)} \,
\sum_{p_2,s_2=0}^\infty
\tilde{A}_{p_2s_2}^{(3,3)}
\nonumber \\
&&\times\,\cos{[2(p_2+s_2)+3]k_2y_2}\,
\frac{1}{2}\{
\frac{\sin{(p_1+s_1+1)v^\prime}}{p_1+s_1+1}\mid_v^{v+\Delta^\prime}
+\frac{\sin{(p_1+s_1)v^\prime}}{p_1+s_1}\mid_v^{v+\Delta^\prime}\} 
\nonumber \\
&+&i\epsilon\,\bar{\chi}_{310}(x)\gamma^{\mu}\,\partial_\mu\chi_{310}(x) 
\,\cos^3{k_2y_2}
\sum_{p_1,s_1=0}^\infty
\tilde{A}_{p_1s_1}^{(3,3)} \,
\sum_{p_2,s_2=0}^\infty
\tilde{A}_{p_2s_2}^{(1,1)}
\nonumber \\
&&\times\,\cos{[2(p_2+s_2)+1]k_2y_2}\,
\frac{1}{2}\{
\frac{\sin{(p_1+s_1+2)v^\prime}}{p_1+s_1+2}\mid_v^{v+\Delta^\prime}
+\frac{\sin{(p_1+s_1+1)v^\prime}}{p_1+s_1+1}\mid_v^{v+\Delta^\prime}\} 
\nonumber \\
&\simeq&i\epsilon\,\bar{\chi}_{130}(x)\gamma^{\mu}\,\partial_\mu\chi_{130}(x) 
\,\cos^3{k_2y_2}
\sum_{p_1,s_1=0}^\infty
\tilde{A}_{p_1s_1}^{(1,1)} \,
\sum_{p_2,s_2=0}^\infty
\tilde{A}_{p_2s_2}^{(3,3)}
\nonumber \\
&&\times\,\cos{[2(p_2+s_2)+3]k_2y_2}\,
\frac{\Delta^\prime}{2}\{
\frac{\cos{(p_1+s_1+1)(k_1y_1+k_3y_3)}}{p_1+s_1+1}
+\frac{\cos{(p_1+s_1)(k_1y_1+k_3y_3)}}{p_1+s_1}\} \nonumber \\
&+&
\,i\epsilon\,\bar{\chi}_{310}(x)\gamma^{\mu}\,\partial_\mu\chi_{310}(x) 
\,\cos^3{k_2y_2}
\sum_{p_1,s_1=0}^\infty
\tilde{A}_{p_1s_1}^{(3,3)} \,
\sum_{p_2,s_2=0}^\infty
\tilde{A}_{p_2s_2}^{(1,1)}
\,\cos{[2(p_2+s_2)+1]k_2y_2}\,
\nonumber \\
&&\times\,\frac{\Delta^\prime}{2}\{
\frac{\cos{(p_1+s_1+2)(k_1y_1+k_3y_3)}}{p_1+s_1+2}
+\frac{\cos{(p_1+s_1+1)(k_1y_1+k_3y_3)}}{p_1+s_1+1}\} 
\label{ap2} 
\end{eqnarray}
where
\begin{eqnarray}
&&\tilde{A}_{p_1s_1}^{(1,1)}=
(f_{4p_1+1}^{(1)*}g_{4s_1+1}^{(1)}+g_{4p_1+1}^{(1)*}f_{4s_1+1}^{(1)}
+f_{4p_1+1}^{(1)*}f_{4s_1+1}^{(1)}
-g_{4p_1+1}^{(1)*}g_{4s_1+1}^{(1)})\label{ap2c} \\
&&\tilde{A}_{p_2s_2}^{(3,3)}=
(f_{4p_2+3}^{(3)*}g_{4s_2+3}^{(3)}+g_{4p_2+1}^{(3)*}f_{4s_2+3}^{(3)}
+f_{4p_2+3}^{(3)*}f_{4s_2+3}^{(3)}
-g_{4p_2+3}^{(3)*}g_{4s_2+3}^{(3)}\,)\label{ap2d} 
\end{eqnarray}
Here $f_n$ and $g_n$'s are constant. Then the total 
effective Lagrangian may be written as
\begin{eqnarray}
{\cal L}^{eff}_{fk2} &=&
\frac{i}{2}\lim_{x^\prime\rightarrow x}
\partial_\mu\;\left(
\,\bar{\chi}_{130}(x^\prime),
\bar{\chi}_{310}(x^\prime)
\bar{\chi}_{330}(x^\prime)
\,\right)\,{\bf \tilde{M}}\,
\gamma^{\mu}
\left( 
\begin{array}{c}
\chi_{130}(x)\\
\chi_{310}(x) \\
\chi_{330}(x)
\end{array}
\right)
\label{ap3} 
\end{eqnarray}
where
\begin{eqnarray}
{\bf \tilde{M}}&=&\left(
\begin{array}{ccc}
\tilde{{\cal A}}&\tilde{{\cal B}}&\tilde{{\cal C}} \\
\tilde{{\cal B}}&
\tilde{{\cal D}}&0 \\
\tilde{{\cal C}}&0&0
\end{array}
\right) \label{ap4} 
\end{eqnarray}
Here
\begin{eqnarray}
\tilde{{\cal A}}&\simeq&
\,\epsilon\cos^3{k_2y_2}
\sum_{p_1,s_1=0}^\infty
\tilde{A}_{p_1s_1}^{(1,1)}\,\tilde{T}_{p_1,s_1}^{(1,3)}(y_1) 
\sum_{p_2,s_2=0}^\infty
\tilde{A}_{p_2s_2}^{(3,3)}
\,\cos{[2(p_2+s_2)+1]k_2y_2} \label{ap4a} \\
\tilde{{\cal B}}&\simeq&
\cos^3{k_2y_2}
\sum_{p,s=0}^\infty A_{ps}^{(1)}(y_2)
\,T_{p,s}(y_1) 
\label{ap4b} \\
\tilde{{\cal C}}&\simeq&
\cos^3{k_2y_2}
\sum_{p,s=0}^\infty A_{ps}^{(3)}(y_2)
\,T_{p,s}(y_1) 
\label{a4c} \\
\tilde{{\cal D}}&\simeq&
\,\epsilon\cos^3{k_2y_2}
\sum_{p_1,s_1=0}^\infty
\tilde{A}_{p_1s_1}^{(3,3)}\,
\tilde{T}_{p_1,s_1}^{(3,1)}(y_1) 
\sum_{p_2,s_2=0}^\infty
\tilde{A}_{p_2s_2}^{(1,1)}
\,\cos{[2(p_2+s_2)+3]k_2y_2} \label{ap4d} 
\end{eqnarray}
where 
\begin{eqnarray}
&&
\tilde{T}_{p_1,s_1}^{(1,3)}(y_1)\,=\,
\frac{\Delta^\prime}{2}\{
\frac{\cos{(p_1+s_1+1)(k_1y_1+k_3y_3)}}{p_1+s_1+1}
+\frac{\cos{(p_1+s_1)(k_1y_1+k_3y_3)}}{p_1+s_1}\} \label{ap4e1} \\
&&\tilde{T}_{p_1,s_1}^{(3,1)}(y_1)\,=\,
\frac{\Delta^\prime}{2}\{\frac{\cos{(p_1+s_1+2)(k_1y_1+k_3y_3)}}{p_1+s_1+2}
+\frac{\cos{(p_1+s_1+1)(k_1y_1+k_3y_3)}}{p_1+s_1+1}\} 
\label{ap4e2} \\
&& T_{p,s}(y_1)\,=\,
\frac{\Delta^\prime}{(p+s)(p+s+1)}[\,
\sin{(p+s)\Delta}\,\cos{(p+s+1)(k_1y_1+k_3y_3)} \nonumber \\
&&+\sin{(p+s+1)\Delta}\,\cos{(p+s)(k_1y_1+k_3y_3)}]
\label{ap4e3}
\end{eqnarray}
Note that $\Delta^\prime<<2\pi$ is employed in (\ref{ap4a}-\ref{ap4d}) 
since the aim is to study the small 
scales point-wise as much as possible (while without 
causing any ambiguity due to the delta function on the brane).
We observe that none of 
the terms in $\tilde{{\cal A}}$, $\tilde{{\cal B}}$,  $\tilde{{\cal C}}$, 
$\tilde{{\cal D}}$  blows up as 
$\Delta\rightarrow\,0$ or $\Delta^\prime\rightarrow\,0$ or
$y_2\rightarrow\,0$ or $y_3\rightarrow\,0$.
In the light of this observation denote the 
maximum possible values of 
$\frac{\tilde{{\cal A}}}{\epsilon}$, 
$\frac{\tilde{{\cal D}}}{\epsilon}$; and the minimum possible values 
of $\tilde{{\cal B}}$, $\tilde{{\cal C}}$ by 
\begin{equation}
(\frac{\tilde{{\cal A}}}{\epsilon})_{max}\,=\,A_{mx}~,~~ 
(\frac{\tilde{{\cal D}}}{\epsilon})_{max}\,=\,D_{mx}~,~~
\tilde{{\cal B}}_{min}\,=\,B_{mn}~,~~ 
\tilde{{\cal C}}_{min}\,=\,C_{mn} \label{ap6}
\end{equation}
It is always possible to choose $\epsilon$ so small (i.e to choose the 
breaking of the symmetry (\ref{a3a}) so small) that
\begin{equation}
X\,\gg\,Y \label{ap7}
\end{equation}
is always satisfied, where $X$ stands for  the smaller of $B_{mn}$, 
$C_{mn}$ and $Y$ stands for the greater of $\epsilon\,A_{mx}$, 
$\epsilon\,D_{mx}$. So
\begin{equation}
\tilde{{\cal A}}, 
\tilde{{\cal D}}\,\ll\, 
\tilde{{\cal B}}, 
\tilde{{\cal C}} \label{ap8}
\end{equation} 
provided that $\epsilon$ is taken sufficiently small.
In other words provided $\epsilon\ll\,1$ is sufficiently small
\begin{eqnarray}
{\bf \tilde{M}}\,\simeq\,
\left(
\begin{array}{ccc}
0&\tilde{{\cal B}}&\tilde{{\cal C}} \\
\tilde{{\cal B}}&0&0 \\
\tilde{{\cal C}}&0&0
\end{array}
\right) 
\label{ap9}
\end{eqnarray}
I take $\epsilon$ such that Eq.(\ref{ap9}) is 
satisfied. Therefore the conclusions about the spectrum of the fields 
at the points $k_1y_1\,\neq\,k_3y_3$ essentially remain the same at the 
points $k_1y_1\,\simeq\,k_3y_3$. In fact it would be enough for us to have 
the relation given in (\ref{ap9}) up to very small length scales in the 
order of Planck scale. In that case the length scales below this scale 
(where the string theory \cite{Hill-Pokorski-Wang} or another quantum 
gravity scheme prevails) would be irrelevant to quantum field theory. 

\section{Pauli-Villars-like regularization}

To compare the propagators in the scales larger and smaller than 
the sizes of the extra dimensions one 
should first put ${\cal L}_{fk2}$ into its canonical form by dividing 
$S_{fk2}$ in (\ref{a9}) by the factor in front of it, $N$ given by
\begin{equation}
N\,=\,
\frac{\epsilon L_1L_2L_3}{4\pi}
(f_{1}^*g_{1}+g_{1}^*f_{1}+f_{1}^*f_{1}-g_{1}^*g_{1})
(f_{3}^{\prime*}g_{3}^\prime+g_{3}^{\prime*}f_{3}^\prime+f_{3}^{\prime*}
f_{3}^\prime-g_{3}^{\prime*}g_{3}^\prime)
\label{a21}
\end{equation}
Note that dividing the total action by an overall constant does not change 
the result because it does not change the equations of motion. In 
general $\chi_{130}$ may have a mass $m$ 
at the scales larger than the sizes of the extra dimensions (e.g. through
a mass term similar to $S_{fk2}$ in from as mentioned before). 
So at the scales larger than the size of the extra dimensions the general 
form of the propagator of $\chi_{130}$ is
\begin{equation}
D(p)\,=\,\frac{i}{\not{p}+m} \label{a22a}
\end{equation}
At the scales 
smaller than the size of extra dimensions the effective propagator 
follows from (\ref{a20}) is the sum of the propagators \cite{DeWitt}
due to $\psi_1$ and $\psi_2$; $D_1$, $D_2$ 
\begin{eqnarray}
D_{eff}(p)&=&D_1(p)\,+\,D_2(p)\sim
\frac{1}{B^\prime}[\frac{i}{\not{p}+m_1}
-\frac{i}{\not{p}+m_2}]\,=\,\frac{i(m_2-m_1)}
{B^\prime(\not{p}+m_1)
(\not{p}+m_2)} \label{a22b} \\
&&B^\prime\,=\,N\,B(y)\cos^3{k_2y_2}\cos{k_3y_3} \nonumber
\end{eqnarray}
where $m_1$, $m_2$, in general, may depend on $y_1$, $y_2$, and I have 
assumed for sake of generality that $\psi_1$, $\psi_2$ may 
have two different effective masses, at scales smaller than the size of 
extra dimensions, that 
may be induced by spin connection terms, Higgs mechanism, or some other 
mechanism. 
Note that the internal fermion lines in a Feynman diagram must be 
identified by (\ref{a22b}) rather than (\ref{a22a}) because the 
off-diagonal terms in (\ref{ap4}) are dominant for all energies (possibly 
at Planck scale or higher scales) as observed in (\ref{ap8}). 
On the other hand the fermion external lines should be 
identified by $\chi_{130}$ or $\chi_{310}$ (that may come from the terms 
of the form, $\chi_{130}X\chi_{130}$, $\chi_{310}X\chi_{310}$, 
$X$=$\not\Omega$, $\phi$ where $\Omega_\mu$, $\phi$ are gauge, scalar 
fields, respectively). For the diagrams containing a single fermion 
internal line this scheme is quite similar to Pauli-Villars 
regularization. 
However this scheme is not wholly equivalent to Pauli-Villars 
regularization in 
the general case \cite{Pauli-Villars,Itzykson-Zuber}. 
For the Feynman diagrams containing a 
single fermion internal line this scheme essentially amounts to 
Pauli-Villars regularization when $m_1\neq m_2$ while it amounts to finite 
renormalization when $m_1=m_2$. 
The implications of this scheme for higher 
number of fermion internal lines and its comparison with Pauli-Villars 
regularization needs further study. In a minimal scheme it needs, at 
least, the incorporation of a $\chi\not\Omega\chi$ 
or $\chi\phi\chi$ type of term into Lagrangian including the study of the 
effects of the higher modes of $\Omega_\mu$ or $\phi$.  This is a quite 
tedious and intricate task and needs a separate study by its own. However 
one may see 
the essential lines of the regularization by (imposing periodic boundary 
conditions for $X$ in all extra dimensions and) considering the zero mode 
of $X$ in $\chi_{130}X\chi_{130}$ and
$\chi_{310}X\chi_{310}$. Such a crude analysis suggests that  
this regularization is rather similar to Pauli-Villars 
regularization. These points need a separate study by its own and 
should be considered in future studies.

\section{Conclusion}

In summary I have introduced a scheme where there is a single usual 
particle $\chi_{130}$ at the scales larger than the sizes of the extra 
dimensions while at the smaller scales the system consists of a particle - 
ghost-like pair given in (\ref{a20}). This effectively amounts to 
Pauli-Villars-like regularization that is essentially equivalent 
to Pauli-Villars regularization whenever a single fermion internal line 
is involved in a Feynman diagram. The situation for higher number of 
fermion internal lines is nontrivial and needs further study. The 
indication that the extra dimensional metric reversal symmetry 
may be relevant to regularization of the divergences of quantum field 
theory in addition to its relevance to the cosmological constant and zero 
point energy problems may suggest that this symmetry may be a fundamental 
symmetry of nature with much deeper implications. 
However this study should be regarded as an example of the possibility of 
inducing a Pauli-Villars like regularization through extra dimensions 
rather than a generic possibility since the realization of this scheme 
needs non-trivial boundary conditions and extra-dimensional symmetries. In 
my opinion a more detailed study of these points, and the extension of 
this study to other fields, such as gauge and scalar fields, and its 
possible relation with Lee-Wick model \cite{Lee-Wick} should be considered 
in future. I do not anticipate extreme difficulty in the extension of this 
scheme to the other fields provided that the discrete symmetries and the 
anti-periodic boundary conditions employed here for some of the extra 
dimensions is used in these extensions as well. All these points need 
separate studies by their own.

\begin{acknowledgments}
This work was supported in part by Scientific and Technical Research 
Council of Turkey under grant no. 107T235.
\end{acknowledgments}



\bibliographystyle{plain}

\end{document}